\documentclass[final,5p,times,twocolumn]{elsarticle}
\usepackage{hyperref}
\usepackage{epstopdf}

\journal{Journal of \LaTeX\ Templates}

\begin{document}

\begin{frontmatter}

\title{Evolution and Recent Developments of the Gaseous Photon Detectors Technologies}

\author{Fulvio Tessarotto\fnref{myfootnote}}
\address{INFN, Sezione di Trieste, Trieste, Italy}
\fntext[myfootnote]{email:fulvio.tessarotto@ts.infn.it}




\begin{abstract}
The evolution and the present status of the gaseous photon detectors technologies are reviewed.
The most recent developments in several branches of the field are described, in particular the installation and commissioning of the first large area MPGD-based detectors of single photons on COMPASS RICH-1.
Investigation of novel detector architectures, different
materials and various applications are reported, and the quest for visible light gaseous photon detectors
is discussed.
The progress on the use of gaseous photon detector related techniques in the
field of cryogenic applications and gaseous or liquid scintillation imaging are presented.

\end{abstract}

\begin{keyword}
Photon detectors \sep gaseous detectors \sep MPGD \sep THGEM \sep CsI \sep photocathodes
\end{keyword}

\end{frontmatter}


\section{Introduction}
\par
The development of gaseous detectors started more
than 100 years ago and provided the fundamental physics research community
with great inventions: the Wire Counter, the Cloud Chamber,
the Geiger-Mueller Counter, the Spark Chamber, the Bubble Chamber,
the Parallel Plate Avalanche Chamber, the Multi-Wire Proportional Chamber (MWPC),
the Drift Chamber, the Time Projection Chamber (TPC), the Limited Streamer Tubes,
the Resistive Plate Chamber and many other detectors allowed to perform critical
and previously impossible measurements. In the last 30 years various
technologies, collectively named Micro-Pattern Gaseous Detectors (MPGDs) have been
introduced to overcome the particle flux capability limitations of MWPCs and achieve
higher space resolution,
starting from the Micro-Strip Gas Counter (MSGC) \cite{Oed-1988-MSGC}, followed by
the successful and wide-spread MicroMegas (MM) \cite{Giomataris-1996-MM} 
and Gas Electron Multiplier (GEM) \cite{Sauli-1997-GEM} detectors.
\par
The use of old gaseous detector technologies is declining but an impressive
development of new techniques, based both on the consolidation and spread of recent ideas
and on the appearance of new elements is opening fascinating perspectives for the years to come.
\par
A similar evolution took place in the domain of photon detection leading to the
introduction of many successful devices and techniques; 
nowadays both vacuum based Photon Detectors (PDs), in particular
Multi-Anode PhotoMultiplier Tubes and Micro-Channel Plates and solid state
based PDs, in particular Si-PMTs and Avalanche Photo-Diodes are commercially
available technologies and provide excellent and greatly
improving performance; their use is increasing and their application range is spreading.
\par
Gaseous PDs instead are not commercially available, but they offer specific
and unique characteristics: a cost-effective solution to cover
very large areas with photosensitive detectors, very low material budget
and minimal sensitivity to magnetic fields.
\par
Gaseous PDs have played an essential role in establishing Ring Imaging CHerenkov
Counters (RICHs) technology and the continuous development of gaseous PDs
has recently been accelerated.
\par
This article aims to illustrate this acceleration, after mentioning some of the
main characteristics and achievements in the field of gaseous PDs,
and to describe some of the present activities
which can potentially lead to important future advancements.

\section{MWPC-based PDs}
\par 
The need from High Energy Physics
Experiments of performing hadron identification in a wide range of momenta and over
a wide angular acceptance motivated the quest for
equipping large area of efficient detectors of single Cherenkov photons 
at affordable cost.
In 1980 suitable photo-ionizing agents: Tri-Ethyl-Amine (TEA)
\cite{Seguinot-1980-TEA} and Tetrakis Dimethylamine Ethylene (TMAE) \cite{Anderson-1980-TMAE}
became available and could be coupled with large area position sensitive gaseous
detectors (MWPCs).
Two conflicting requirements however had to be composed: a high efficiency in conversion and
detection of Cherenkov UV photons, implying a multiplication gain large enough to allow
detecting single photoelectrons and a limited sensitivity to the UV photons copiously emitted
in the large avalanche process without using a photon quencher which would kill
the primary signal.
\par
The first solution was the Multi-step Avalanche Chamber, developed by Charpak and Sauli
\cite{Charpak_Sauli_1978_Multi-step} and used by the RICH of E605 Experiment
\cite{Adams-1983-E605} \cite{McCarty-1986-E605} at FNAL with TEA vapour and CaF$_2$ windows.
MWPC-based PDs with TEA and CaF$_2$ windows
were also used by CLEO III \cite{Artuso-2003-CLEO}.
Using TEA in a CH$_4$ atmosphere, the detected photon
wavelength range is limited to the far UV interval between 135 and 165 nm, where
very few materials are transparent and the chromatic dispersion is large: E605 indeed
used He, the lowest chromaticity gas, as radiator.
\par
TMAE has a low photo-ionization threshold (5.3 eV) and is compatible with the
use of fused silica windows: the resulting effective wavelength range for photon
detection is between 160 nm and 220 nm.
The largest RICH detectors ever built: the DELPHI RICHs \cite{Albrecht-1999-DELPHI},
the SLD CRID \cite{Vavra-1999-SLD-CRID} and the OMEGA RICH \cite{Siebert-1999-OMEGA}
\cite{Mueller-1996-OMEGA}
used TMAE and a PD based on the TPC RICH concept \cite{Apsimon-1986-OMEGA}.
Due to the low TMAE vapor pressure at room temperature the photon conversion
gap needs to be centimeters thick and the detectors were operated at high temperature
(40 degrees or even 50 degrees in the case of CERES RICH \cite{Baur-1994-CERES}).
The photoelectrons were drifted in a uniform electric field toward MWPCs to obtain a 3D
imaging, which was needed to correct for the parallax error
affecting the photon conversion in a thick gap. The long drift time made these detectors
intrinsically slow and spurious signals were generated by the conversion of photons
produced in the amplification process; against them special blinds had been inserted between
wires in the RICH MWPCs \cite{Anassontzis-1992-DELPHI-BARREL}.
The extreme chemical reactivity of TMAE severely restricts the choice of detector materials
and the tolerance of gas impurities \cite{Hallewell-1994-TMAE}.
\par
TMAE emits visible light: TMAE-based optical readout of RICH detectors
\cite{Breskin-1988-optical-readout} was implemented in the NA35 Experiment
\cite{Baechler-1994-NA35}.
\par
The photo-converting property of a thin film of CsI for UV photons is known
since 1956 \cite{Philipp-Taft-1956-CsI}, and vacuum operated photocathodes had been
developed for space astronomy \cite{Carruthers-1969-CsI}.
The first position sensitive gas detectors with CsI photocathodes were developed
at the end of the 80s \cite{Charpak-1969-CsI} \cite{Seguinot-1990-CsI}.
\par
Semi-transparent CsI photocathodes are commonly used in vacuum-based PDs, and they could
as well be used for gaseous-based PDs. Their photon conversion efficiency is limited by
the need to deposit a thin conductive layer on the window which absorbs a non-negligible
fraction of UV photons
and by the need to optimize the thickness of the CsI layer \cite{Lu-1994-CsI}:
a thinner layer has lower photon
absorption probability, a thicker one has a lower escape probability for the photoelectron.
The strict requirement on CsI thickness uniformity makes the production of large area
semitransparent CsI photocathodes technically difficult. The reflective option has indeed
been chosen for all CsI gaseous-based PD photocathodes used in large experiments.
\par
Due to its sensitivity to the substrate and its degradation in the presence of humidity
many years were needed to develop an optimized procedure for production and handling of
large area CsI photocathodes for gaseous PDs. This technology was pioneered by T. Ypsilantis
ad J. S\'eguinot \cite{Seguinot-1996-CsI-RICH} and developed by the RD26 Collaboration
\cite{Piuz-1999-RD26} coordinated by F. Piuz. 
The technology of large area gaseous PDs with solid photo-converter 
for RICH applications represented a real breakthrough; it is based on the use of
MWPCs with thin gap (2 mm), cathode wires of 100 $\mu$m diameter and 2 mm pitch, anode 
wires of 20 $\mu$m diameter and 4 mm pitch and CsI photocathodes segmented in 8 $\times$ 8
mm$^2$ pads. \cite{DiMauro-1999-RD26}. The PDs are operated in pure CH$_4$
to minimize the photoelectron backscattering \cite{DiMauro-1996-CsI-CH4}.

\par
An optimized photocathode production procedure \cite{Braem-2003-CsI}
\cite{Hoedlmoser-2006-CsI} for achieving high QE was defined, involving
proper substrate preparation (Cu-clad PCB coated with 7 $\mu$m Ni and 0.5 $\mu$m Au),
ultrasonic bath cleaning and out-gassing, a slow deposition ($\sim$1 nm/s) of a 300 nm thick
CsI layer by thermal evaporation (at about 10$^{-7}$ mbar) followed by a thermal
treatment (8 h at 60 $^\circ$C). 
Essential parts of the procedure are: the mapping of the local photocathode response,
the encapsulation and flushing with dry gas during storage and the handling under
controlled atmosphere using glove-boxes.

\par
RD26-type MWPC-based PDs have been \cite{Piuz-2003-RICH} and still are
successfully used by several experiments: NA44 \cite{Fabjan-1995-TIC} at CERN,
HADES \cite{Zeitelhack-1999-HADES} at GSI, COMPASS \cite{Tessarotto-2014-RICH} 
at CERN (5.5 m$^2$ active area), STAR \cite{Braem-2003-STAR} at RHIC,
HALL-A \cite{Iodice-2005-HALL-A}
at JLAB,  ALICE \cite{DiMauro-2005-ALICE} at CERN (10 m$^2$ active area), etc.
The impact of MWPC-based PDs on the physics results of the experiments is very
relevant and they can be considered the classical successful gaseous PDs.
Their use had been proposed for the upgrade of ALICE HMPID
\cite{Acconcia-2014-C4F8O-VHMPID}.
\par
The aging of CsI has been investigated in detail \cite{Hoedlmoser-2007-QE} and is
reported to result in a severe decrease of the quantum efficiency after a total
collected charge of few mC/cm$^2$. Aging is mainly caused by the Ion Back-Flow (IBF)
to the photocathode, namely the bombardment of the CsI layer by the positive ions
generated in the multiplication process.
\par
The CsI photocathodes used in gaseous PDs
preserve their QE over a
period of many years with at most a modest decrease \cite{DeCataldo-2014-aging}
\cite{Sozzi-2014-aging} \cite{DallaTorre-2014-QE}.
\par
MWPC-based PDs have intrinsic limits on rate capability and time resolution;
they have to operate at relatively low gain, because of their open geometry: secondary
effects from photon feedback and large ionization events can otherwise generate
electrical instabilities \cite{DallaTorre-2005-MWPC-CsI}.
Their readout signals derive from the slow motion of ions and they have long detector memory,
even when coupled to upgraded front-end electronics \cite{Neyret-2006-RICH-APV}.
The aging is accelerated by their large IBF.
To achieve the higher performance needed in high rate applications, a technology
change was needed \cite{DallaTorre-2011-Status-PD}.

\section{GEM-based PDs}

\par
The progress of photo-lithography and other technologies in mastering the production of
complex patterns with accuracy at the 10 $\mu$m level or better allowed the
development of many MPGDs
(MSGCs \cite{Oed-1988-MSGC},
Micro-Gap \cite{Angelini-1993-Micro-Gap},
Micro-Dot \cite{Biagi-1995-Micro-Dot},
Compteur \`a Trous \cite{Bartol-1996-CAT},
MM \cite{Giomataris-1996-MM},
GEM \cite{Sauli-1997-GEM},
Micro-Groove \cite{Bellazzini-1998-Micro-Groove},
Micro-Wire Chamber \cite{Adeva-1999-Micro-Wire},
Micro-Pin Array \cite{Rehak-2000-Micro-Pin}
Micro-Pixel \cite{Ochi-2002-Micro-Pixel},
Field Gradient Lattice \cite{Dick-2004-Field-Gradient}, etc.).
Among them MM and GEM are fully mature
and consolidated technologies, ubiquitously used for the detection of particles.
\par
MMs are parallel plate chambers built with a
thin metal grid stretched and kept at a small, uniform distance (about 100 $\mu$m)
from the readout electrode by
isolating supports; operated with electric field above 30 kV/cm they provide
proportional multiplication of the electric charge entering the high field
region from the open spaces of the grid.
MMs are robust and have low sensitivity to small gap variation or imperfections.
They provide
good energy resolution and, thanks to the large difference in the electric
field in the regions above and below the grid, a dominant fraction of the ions
produced in the multiplication is collected by the grid \cite{Colas-2004-MM-IBF}.
\par
GEM consist of 50 $\mu$m thin, copper-clad polyimide foils with a high density
regular matrix of holes, produced by photo-lithography and chemical etching processes.
Inserted between a
drift and a collection electrode and properly biased, they provide proportional
multiplication of the ionization electrons produced in the drift region and efficient
transfer of electrons into the collection region. The multiplication occurs mainly inside
the holes of the GEM, the chamber readout electrode is physically independent
from the GEM and the signal is fast, being generated only by electrons. 
Several GEM layers can be cascaded in the same
detector \cite{Bressan-1999-GEM}, allowing discharge free operation, good
space and time resolution \cite{Ketzer-2004-GEM} \cite{Morman-2003-GEM-UV}
and high rate capability \cite{Benlloch-1998-GEM}.
\par
Detectors based on a triple-GEM amplification have been
pioneered by the COMPASS experiment at CERN \cite{Altunbas-2002-GEM}, and are
routinely used in particle physics experiments (LHCb \cite{Bencivenni-2002-LHCb-GEM},
PHENIX \cite{Fraenkel-2005-PHENIX-GEM}, KLOE-2 \cite{Bencivenni-2007-KLOE-GEM}
TOTEM \cite{Bagliesi-2010-TOTEM-GEM}) and proposed for the upgrade of
CMS \cite{Abbaneo-2010-CMS-GEM} \cite{Abbaneo-2015-CMS-GEM} \cite{Abbaneo-2017-CMS-GEM}.
\par
The MPGD developers community is actively improving the MM, GEM and other technologies;
it has a reference, world-wide forum of discussion,
where common standards are defined and access to common facilities is granted,
in the RD51 Collaboration \cite{RD51-proposal} at CERN.
Innovative proposals and freshly obtained test
results are frequently compared and new ideas continuously appear and compete
in this forum.
\par
The use of a GEM as a first amplifier in a gaseous PD \cite{Chechick-1998-GEM}
was proposed as soon as the GEM was invented, to exploit its low optical transparency.
After a first prototype with a semi-transparent photocathode \cite{Buzulutskov-2000-GEM-CsI},
a multi-layer GEM structure having a CsI coating
on the top of the first GEM \cite{Morman-2003-GEM-CsI} was developed and successfully operated
by PHENIX-HBD \cite{Anderson-2011-HBD}.
This detector represents the first application of MPGD-based photon counters in an
experiment. The PDs \cite{Kozlov-2004-GEM-HBD} are triple GEMs covering about
1.5 m$^2$. The detector is windowless and operates with a gain of about 5000 in pure
CF$_4$. The CsI layer on the first GEM converts the Cherenkov photons produced in
the 50 cm thick radiator gas volume, over a wide wavelength range (108-200 nm).
The large (6.2 cm$^2$) readout pads provide collective signals from several
(5 to 10) Cherenkov photons.
The identification of low-mass di-leptons requires efficient hadron blindness
which is achieved by using CF$_4$ as radiator medium (so that hadrons in the momentum
range studied by PHENIX do not emit Cherenkov photons)
and a slightly reversed field in the radiator region which
repels the ionization electrons released by charged particle tracks.
\par
GEM-based PDs provide fast signals (time resolution $<$ 2 ns)
\cite{Morman-2003-GEM-CsI} and do suppress the photon feedback thanks to
their closed geometry, but in
order to efficiently detect single photons they should operate at higher
gain than in the PHENIX-HBD case and to be suitable for high rate operation they
should limit the IBF to the photocathode, to preserve it
from aging and avoid secondary effects causing electrical instability.
\par
IBF in multi-GEM structures with semi-transparent \cite{Bondar-2003-GEM-IBF}
or reflective \cite{Morman-2004-GEM-IBF} photocathodes can be reduced to 2\% and 10\%
respectively without affecting the total gain; less than 1\% IBF is obtained with very
low values of the drift field \cite{Breskin-2002-GEM-IBF}; this option however is not
suited for a PD with reflective photocathode because the CsI photoelectron extraction
efficiency at low electric field values at the CsI surface is too low.
\par
More elaborate GEM structures have been introduced to achieve higher IBF suppression.
The Micro Hole and Strip Plate (MHSP) \cite{Veloso-2000-MHSP} has one of the two
GEM surfaces segmented and equipped with microstrips allowing for further amplification or for
ion trapping \cite{Maia-2004-GEM-MHSP}: an IBF of 3$\times$10$^{-4}$ at a gain of 10$^5$ in
Ar/CH$_4$ 95/5 has been obtained using multiple MHSPs in reverse bias mode
\cite{Lyashenko-2007-MHSP}. Extremely low values of the IBF ($\le$ 10$^{-5})$
have been achieved using specially shaped MHSP called COBRA \cite{Lyashenko-2009-COBRA}.
The possibility to collect a large fraction of the ions in the intermediate electrodes
is of particular interest for visible light gaseous PDs
and for TPC applications \cite{Sauli-2006-GEM-IBF} \cite{Berger-2017-ungated-TPC};
for the latter case all GEM electrodes can be used for IBF suppression. 
An optimization of GEM detectors for the ALICE-TPC upgrade
\cite{Ball-2014-ALICE-GEM-IBF} achieved less than 1\% IBF 
in a quadruple GEM structure with largely unbalanced transfer field values and different
hole patterns in different GEM layers \cite{ALICE-TDR-016-ADD-1}.
GEM-based PDs have been proposed for cryogenic applications too.
\par
The typical gain of multi-GEM detectors used as trackers in experiments is mostly
below or around 10$^4$, despite the fact that they achieve 10$^5$ or more in laboratory
tests.
\par
A high-gain alternative electron multiplier has been introduced for
the detection of single photons.

\section{THGEM-based PDs and the Hybrid PD's of COMPASS RICH-1}

\par 
Thick-GEMs (THGEMs) are gaseous electron
multipliers derived from the GEM~design, scaling the geometrical parameters and
changing the production technology: standard Printed Circuit Boards (PCBs) are used
instead of the copper-clad polyimide foils and the holes are obtained by drilling.
They were introduced in parallel by several groups \cite{THGEMS-1} \cite{THGEMS-2}
\cite{Ostling-2003-THGEMS} \cite{THGEMS-3} \cite{THGEMS-4}.
\par
THGEMs are simple, mechanically stiff, electrically robust and cost effective;
they have typical thickness of 0.2-1.2 mm, cylindrical holes with diameter in
the 0.2-1.0 mm range and pitch of 0.4-2.0 mm. Their holes can be provided with a rim,
a clearance ring in the Cu layers around the holes which can vary from 0.0 to 0.2 mm in width.
They can be industrially manufactured in large series and large size using standard
PCB drilling and etching techniques.
\par
In comparison with the GEM case, the space resolution provided by THGEM-based detectors
is modest \cite{Cortesi-2007-THGEM-imaging} ($\sim$1 mm) and the material budget
is large, but the electron collection and transport is more
effective \cite{Chechik-2004-THGEM-UV} (for THGEMs the electron transverse diffusion
is smaller than the hole diameter) and the
achievable gains are larger \cite{Shalem-2006-THGEM-atm} \cite{Shalem-2006-THGEM-low-p}.
\par
THGEMs with different geometrical parameters have been extensively characterized and
their role as electron multipliers and as reflective photocathode has been studied in
detail \cite{Breskin-2007-THGEM-review} \cite{Dallatorre-2010-THGEM-photocurrent}
\cite{Dallatorre-2012-THGEM-detection} \cite{Rocco-2010-PhD-thesis}
\cite{Hamar-2017-Leopard}.
A special role is played by the rim: the maximum achievable gain increases exponentially with
the rim size \cite{Breskin-2007-THGEM-review} but large gain variations over time and significant
gain dependence on the irradiation history \cite{Tessarotto-2009-quest}
\cite{Dallatorre-2014-gain} are seen for large rim THGEMs.
\par
Apart from standard THGEMs, produced from PCB material (FR4), special THGEMs have been produced
using different substrate materials, including noble substrates
(ceramic \cite{Xie-2013-THGEM},
glass \cite{Takahashi-2013-glass-GEM} \cite{Fujiwara-2014-glass-GEM} \cite{Mitsuya-2015-glass-GEM}),
organic substrates (Kevlar, PTFE \cite{Xie-2013-PTFE}, etc.) or using different
production procedures, including water jet or laser drilling and chemical etching.
The technology for Capillary Plates (CP) production, typically used in vacuum-based PDs,
provides multipliers for hole-type MPGDs which could be included in the THGEM category too.
\par
THGEMs can be treated to have the holes covered by highly isolating films (polyurethane)
or the electrodes covered by resistive layers:
an intense investigation of this design, called RETGEM
(PCBs covered with resistive kapton layers) has been performed \cite{Oliveira-2007-RETGEM}.
\par
Interesting THGEMs with different structures have been produced, in particular
the analogue of the GEM COBRA: the THCOBRA \cite{Amaro-2010-THCOBRA},
discussed in Section 5, the Multi-layer THGEM \cite{Cortesi-2017-M-THGEM}
and the Blind-THGEM, also called WELL \cite{Arazi-2012-WELL} (from the name
of a MPGD proposed in 1999 by Bellazzini \cite{Bellazzini-1999-WELL})
namely a THGEM with a closed bottom anode, also in the resistive-anode version,
discussed in Section 5 too.
\par
Chambers hosting multilayer standard THGEMs arrangements 
with CsI~coating on the top of the first THGEM \cite{Breskin-2007-THGEM-review}
have been built and operated in Ar-based and in Ne-based gas mixtures
\cite{Peskov-2010-THGEM-Ne} \cite{Peskov-2012-CsI-THGEM}.
\par
High gain, stable operation in laboratory and at test beam lines \cite{Tessarotto-2013-VCI}
have been reported for small-size prototypes of various configurations,
in particular with triple identical THGEMs, the first one being CsI-coated:
effective gain in the range of~10$^5$-10$^6$ \cite{Tessarotto-2010-development}
are commonly achieved, and time resolutions below 10~ns \cite{Levorato-2012-testbeam}
are typical.
\par
Obtaining the same performance in terms of gain and stability with large or medium
size (300 $\times$ 300 cm$^2$) triple THGEM PDs was more challenging:
a dedicated R\&D program \cite{Dallatorre-2012-THGEM-detection}
\cite{Tessarotto-2013-status-progress}
investigating the origin of non-uniformity of the detector
response and the spark rates as well as the performance of different PD configurations
provided a specific procedure for large area THGEM production
\cite{Levorato-2014-progress-THGEM}, quality assessment and
configuration optimization. 
An investigation of the IBF \cite{Dallatorre-2013-IBF}, which in a standard
triple THGEM configuration approaches 30\%, showed that it can be reduced by
a complete misalignment of the holes in different layers and using unbalanced values
of the electric field in the transfer regions between THGEMs (as in the GEM case
\cite{Sauli-2006-GEM-IBF}).
\par
To achieve a further IBF suppression an alternative architecture
combining MM and THGEM technologies was tested and provided better
results in terms of PD performance and stability for large area prototypes 
\cite{Santos-2016-THGEM-Hybrid-test}.
This hybrid configuration has been chosen for the COMPASS RICH-1 upgrade
\cite{Tessarotto-MPGD2015-hybrid-PDs}, performed in 2016: 
four new MPGD-based PDs \cite{Dasgupta-2017-PhD-thesis}, 
covering a total active area of 1.4 $m^2$ have been installed
replacing MWPC-based PDs which were in operation since 2002;
for the first time MPGD-based detectors of single photons were used in a
running experiment \cite{Levorato-RICH2016-hybrid-PDs}.
\par
\begin{figure}[!htb]
\centering
\includegraphics[scale=1.1]{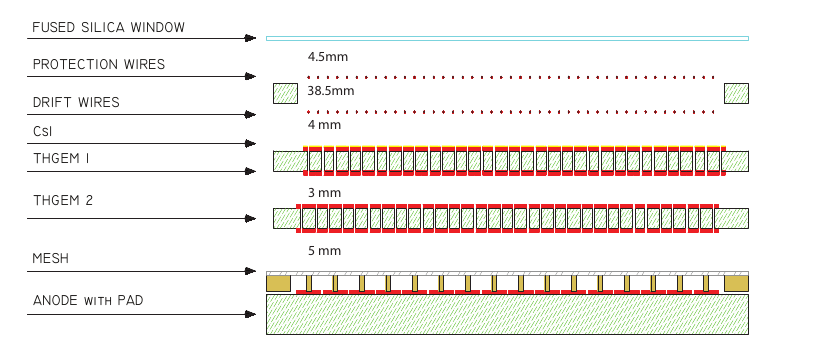}
\caption{Sketch of the hybrid single photon detector: two THGEM layers are coupled to a bulk Micromegas.
The drift wires and the protection wires are shown. The distances between electrodes and to the window
are indicated too. The image is not to scale.}
\label{fig:hybrid}
\end{figure}

The new detector architecture 
consists in a hybrid MPGD combination (see Fig.\ref{fig:hybrid})
of two THGEMs, the first acting as reflective photocathode, and a Micromegas
on a pad segmented anode.

\par
\begin{figure}[!htb]
\centering
\includegraphics[scale=0.31]{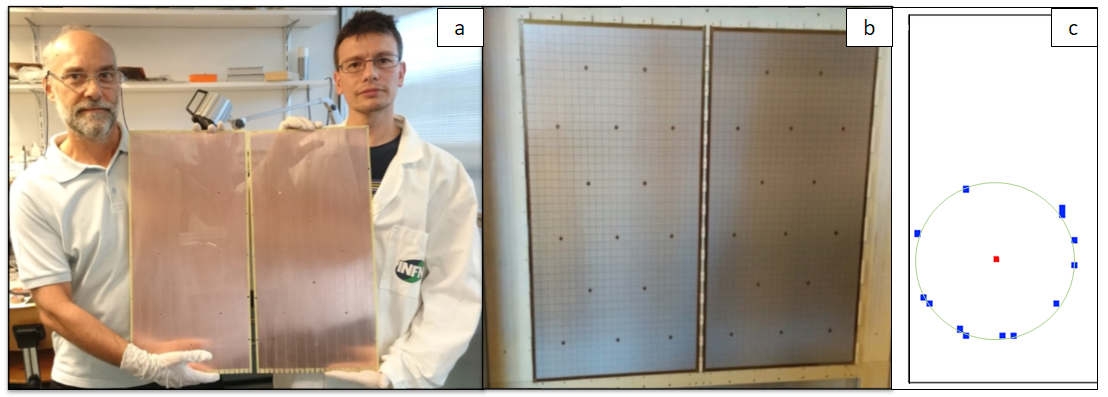}
\caption{(Two COMPASS RICH-1 THGEMs (a) and MMs (b):
they are mounted side by side and cover an active area of $\sim$ 600 $\times$ 600 mm$^2$.
An example of a Cherenkov ring from the new hybrid detectors (c): the diameter is $\sim$ 30 cm.}
\label{fig:MM-THGEM}
\end{figure}

Each of the four PDs
covers a 600 $\times$ 600 mm$^2$ active area and is formed by two identical modules
(600 $\times$ 300 mm$^2$), arranged side by side.
Two planes of wires are
located respectively at 4 mm from the CsI coated THGEM and at 4.5 mm from the fused
silica window of the PD. They allow the tuning of the electric field in the region
above the photocathode in order to optimize the photoelectron collection and repel
the ionization electrons.
All THGEMs (Fig. \ref{fig:MM-THGEM}, a) have the same geometrical parameters:
they are 470 $\mu$m thick (400 $\mu$m dielectric and
2 $\times$ 35 $\mu$m Cu), their holes have 400 $\mu$m diameter,
800 $\mu$m pitch and no rim. Holes located along the external borders have
500 $\mu$m diameter. The electrodes are segmented in 24 mm wide strips and the
voltage bias is individually provided to each sector of the THGEMs
through protection resistors;
accurate raw material selection, post-production treatments and validation tests
were performed.
\par
The two THGEM layers are mounted at a distance of 3 mm, in a configuration of
complete hole misalignment, to achieve the maximum charge spread; 5 mm separate
the middle THGEM from the MMs.
The MMs (Fig. \ref{fig:MM-THGEM}, b) were produced at CERN using the bulk technology
procedure \cite{Giomataris-2006-bulk-MM} \cite{Bouchez-2007-bulk-MM-T2K}: they
have a 128 $\mu$m gap, 18 $\mu$m woven stainless steel wire
mesh with 63 $\mu$m pitch; a square array of 300 $\mu$m diameter pillars with
2 mm pitch keeps the micromesh in place, on a PCB specifically designed for
COMPASS RICH-1.
The square anode pads facing the micromesh have 8 mm pitch and 0.5 mm inter-pad
clearance and are biased at positive voltage; the micromesh, which is the only
non-segmented electrode, is kept at ground potential.
Each anode pad receives the biasing voltage via an individual 470 M$\Omega$ resistor.
\par
The signal is transmitted from the anode pad via capacitive coupling to a 
readout pad facing it, buried inside the anode PCB (at 70 $\mu$m distance from
the anode pad) and connected by a short path to the front-end board connector.
The resistive-capacitive pad scheme dumps the effects of discharges and protects
the front-end
electronics \cite{Neyret-2006-RICH-APV}, which is based on the APV25 chip.

\par
COMPASS hybrid PDs operate with an Ar/CH$_4$ 50/50 gas mixture, at transfer
field values of 1.5 kV/cm between THGEMs and  0.8 kV/cm between the MM and the
THGEMs. The typical effective gain values are few 10$^4$ and the noise level
is $\sim$ 900 equivalent electrons r.m.s. \cite{Dasgupta-2017-PhD-thesis}.
The IBF to the photocathode is $\sim$ 3\%.
A high voltage monitoring program stabilizes the gain by tuning the biases applied
to MM and THGEMs to compensate for the environmental changes of temperature and pressure.

\par
The COMPASS hybrid PDs have been commissioned in 2016 and provided stable
performance during the 2017 run: an example of their
Cherenkov rings is presented in Fig. \ref{fig:MM-THGEM}, c.
The validity of the MPGD-based PD approach for RICH applications is confirmed by
the successful operation of COMPASS hybrid PDs.

\section{Other gaseous PDs sensitive in the UV domain}

Some gaseous PDs could provide unique characteristics for special applications:
very high space resolution, extremely reduced material budget, precise timing etc.
Few examples are mentioned in this section.
\par
Following the idea of active pixel readout of GEMs \cite{Bellazzini-2004-GEM-pixel-ASIC}
a UV-sensitive gaseous detector \cite{Bellazzini-2007-GEM-CsI-CMOS},
based on a CsI semi-transparent photocathode, a GEM charge multiplier and a
self-triggering CMOS analog pixel chip with 105 k pixels at 50 $\mu$m pitch was built and
shown to achieve an accuracy of 4 $\mu$m rms in the coordinates of each GEM avalanche.
\par
\begin{figure}[!htb]
\centering
\includegraphics[scale=0.22]{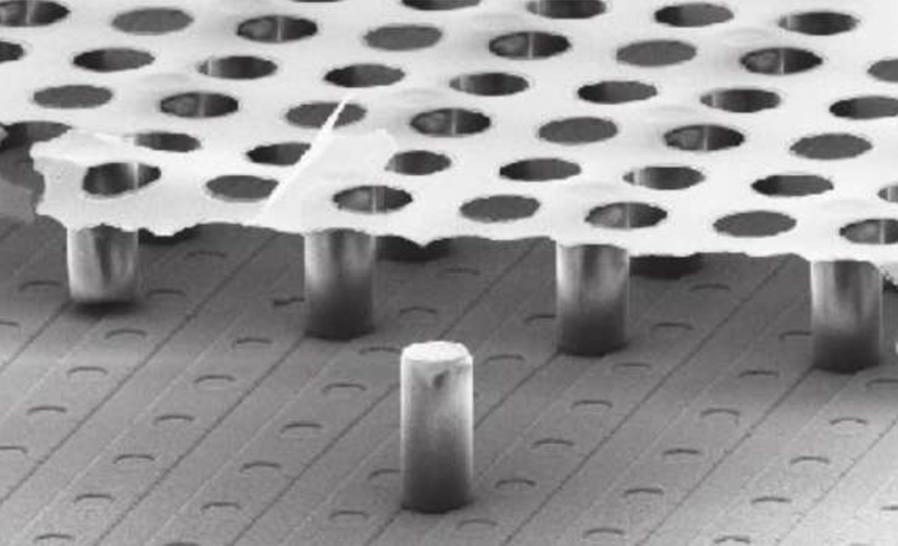}
\includegraphics[scale=0.23]{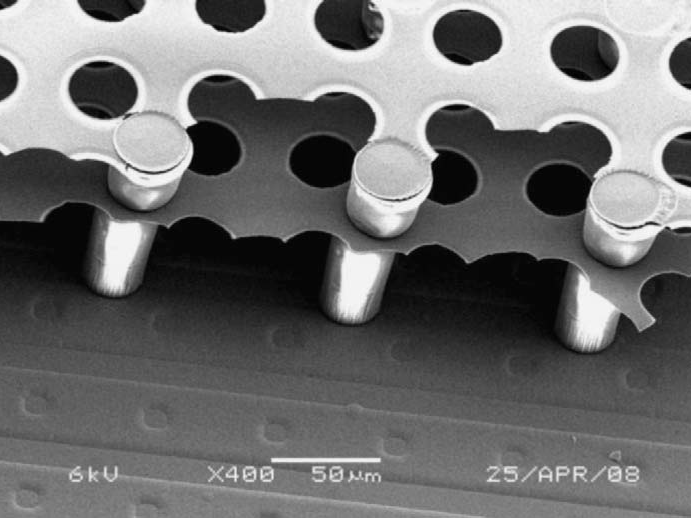}
\caption{Microscope photos of an InGrid (left) and a TwinGrid (right) detector.}
\label{fig:ingrid}
\end{figure}

 
A similar hybrid architecture able to provide excellent space resolution for single
UV photon detection is based on the use of InGrid technology \cite{Chefdeville-2006-InGrid}
consisting of a MM (Al micro-grid) directly integrated by wafer post-processing
production onto a CMOS pixel detector with the complete readout system
(see Fig.\ref{fig:ingrid}).
The micro-grid is sustained on pillars located at the
intersection of four adjacent pixels while the array of micro-grid round holes corresponds
to the array of CMOS pixel centers (square pattern with $\sim$ 55 $\mu$m pitch).
CsI coating of the micro-grid, which has an optical transparency of $\sim$ 20\%,
provides photo-sensitivity. Using a He/i-C$_4$H$_{10}$ 80/20 gas mixture a gain of
6$\times$10$^4$ has been obtained.
Several other detectors with similar architectures (produced by wafer
post-processing) have been tested, including a TwinGrid \cite{Bilevych-2009-TwinGrid}
with a double metal grid onto the CMOS pixel chip.
The use of a 10 $\mu$m thick protection layer of hydrogenated amorphous silicon or
silicon-rich nitride with high resistivity is shown \cite{Bilevych-2011-resistive-Grid}
to quench the sparks and limit the maximum charge provided by discharge signals.
Detectors based on the combination of InGrids with the Timepix ASIC \cite{Llopart-2007-Timepix},
called GridPix \cite{Kaminski-2017-GridPix} have also been used as TPC readout elements and
for X-Ray detection (in the CAST experiment at CERN).
\par
An option to minimize the material budget of a PD is the use of GEMs with Al clad
or thinner Cu clad or even with the Cu etched away
(leaving only the 100 nm thik Cr layer normally used as a tie coat for the
adhesion of Cu): a reduction down to
$\sim$10$^{-3}$ X$_0$ \cite{DuartePinto-2011-thin-GEM}
\cite{Mindur-2017-thin-GEM} could be reached.
\par
Windowless GEMs with CsI are proposed for particle identification at a future
Electron Ion Collider \cite{Accardi-2015-EIC} facility for a focusing RICH and a
for a combined TPC/Cherenkov detector, both operating with pure CF$_4$ or CF$_4$-rich
gas mixtures.
For the focusing RICH a prototype with 1 m long CF$_4$
radiator, a mirror with high reflectivity at $\sim$120 nm and
windowless quintuple-GEMs + CsI, has recently been tested \cite{Blatnik-2015-EIC}.
It is proposed to operate the central tracking TPC with CF$_4$ and equip it 
with gaseous PDs located inside the TPC gas volume, performing UV photon
conversion and electron multiplication in the same gas; these PDs could be similar
to PHENIX HBD GEM-based ones and
could use Cherenkov light signals to identify electrons with the PHENIX HBD technique.
The compatibility of these PDs with the delicate TPC operation is being
actively investigated \cite{Woody-2015-Cherenkov-TPC}.

\par
The THCOBRA \cite{Amaro-2010-THCOBRA} is a THGEM having one of the faces equipped with
additional anode strips winding between circular cathode strips.
Primary avalanches occurring within the holes are
followed by additional ones at the anode-strips vicinity.
\begin{figure}[!htb]
	\centering
	\includegraphics[scale=0.71]{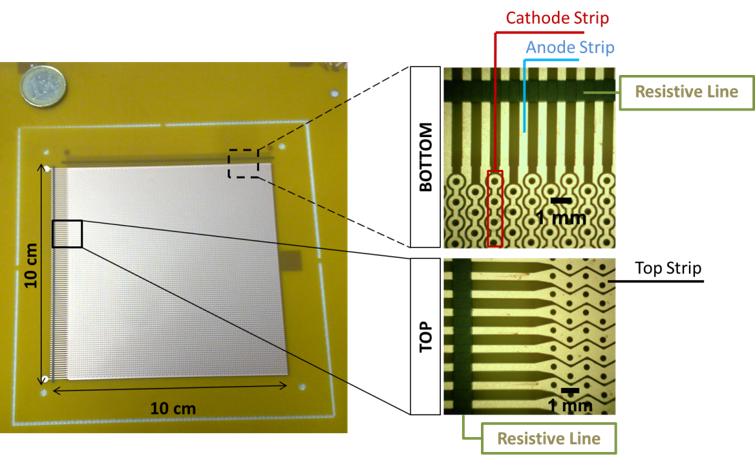}
	\caption{The THCOBRA and its strips.}
	\label{fig:THCOBRA}
\end{figure}
\begin{figure}[!htb]
	\centering
	\includegraphics[scale=0.5]{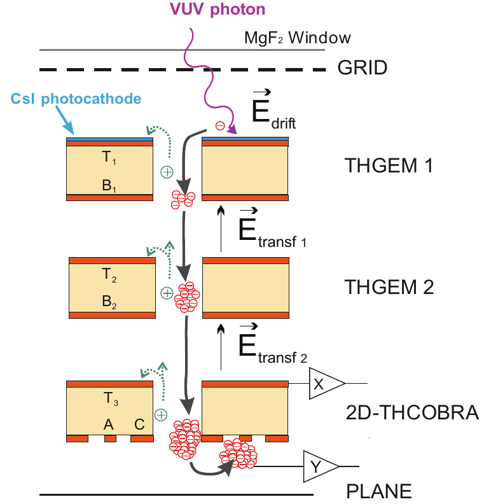}
	\includegraphics[scale=0.5]{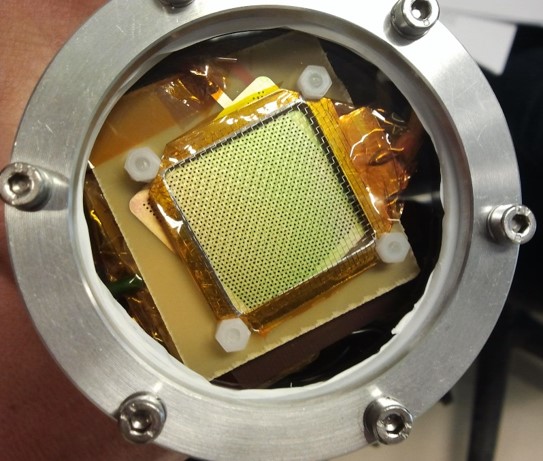}
	\caption{The VUV-THCOBRA PD: the principle scheme (left) and the
		CsI THGEM + THCOBRA PD (right).}
	\label{fig:VUV-THCOBRA}
\end{figure}

The THCOBRA has been used coupled with THGEMs for IBF suppression \cite{Veloso-2011-THCOBRA}.
An evolution of the THCOBRA concept, namely a THCOBRA having the other face segmented in
strips orthogonal to the COBRA strips (see Fig. \ref{fig:VUV-THCOBRA}) has been used as
both electron multiplier and readout element, allowing for a position sensitive detector
to be built without a separate anode readout systems.
A small PD consisting of a THGEM with CsI, a second THGEM and the strip THCOBRA
(see Fig. \ref{fig:THCOBRA} and \ref{fig:VUV-THCOBRA})
has been built and demonstrated to provide good VUV sensitivity and space resolution
with very simple electronics readout.

\par
\begin{figure}[!htb]
	\centering
	\includegraphics[scale=0.3]{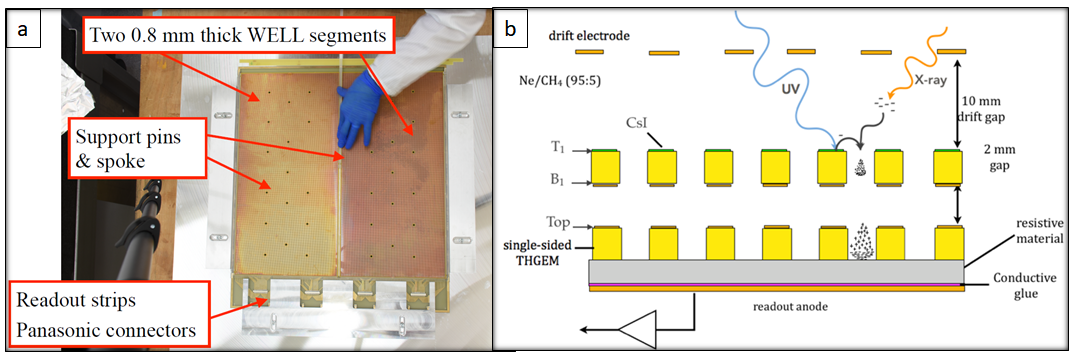}
	\caption{a: THGEMs and b: scheme of the RPWELL PD}
	\label{fig:RPWELL}
\end{figure}

The Resistive-Plate WELL (RPWELL) \cite{Rubin-2013-RPWELL} is a novel gaseous
multiplier based on the WELL concept. It is made of a single-sided copper-clad THGEM electrode,
coupled to a segmented readout anode (pads or strips) through a
thin high bulk-resistivity plate.
It demonstrated discharge-free operation at high gas-avalanche gains and
over a broad ionization range \cite{Moleri-2016-RPWELL} \cite{Bressler-2016-RPWELL}.
It has recently been proposed to exploit the high stability of the RPWELL multiplier
to develop a gaseous PD similar to the COMPASS RICH Hybrid one, with Micromegas
replaced by RPWELLs (see Fig. \ref{fig:RPWELL}), making use of silicate glass resistive plates
with 10$^9$ $\Omega$ cm \cite{Wang-2010-slica-glass}.

\par
\begin{figure}[!htb]
	\centering	
	\includegraphics[scale=0.15]{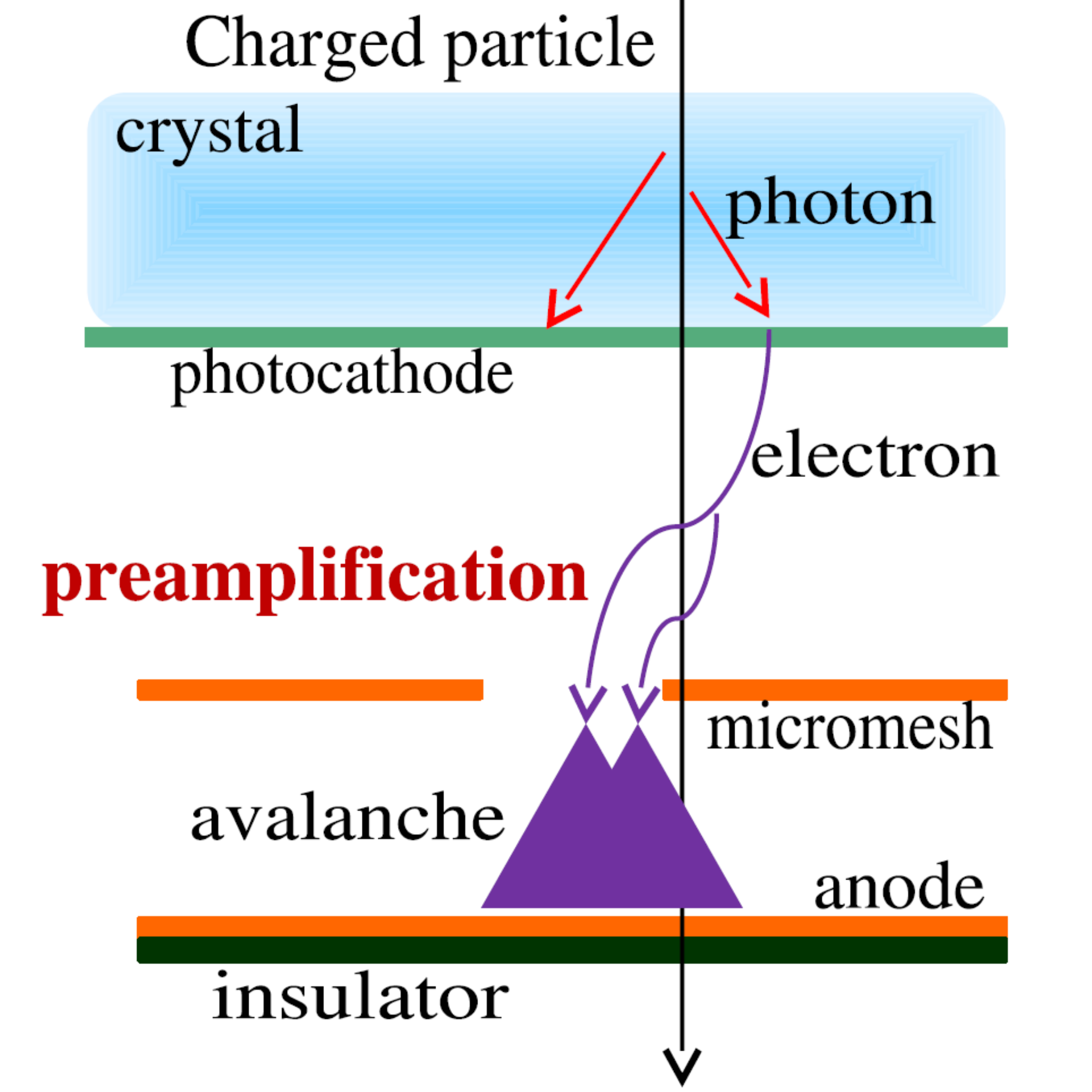}
	\includegraphics[scale=0.35]{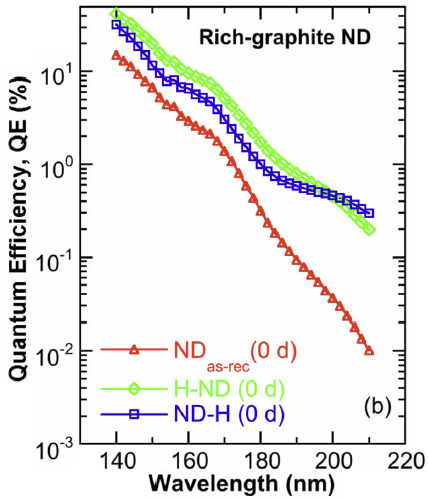}
	\caption{The scheme of the Micromegas inspired high time resolution detector}
	(left) and measured QE of hydrogenated rich-graphite
	nanodiamond grains (right).
	\label{fig:picosec-CsI}
\end{figure}

The quest for a detector able to provide very high time resolution in a high rate environment
triggered the investigation of a Cherenkov light detector with minimal time jitter
\cite{White-2014-picosecond}.
A Micromegas PD has been constructed in order to study the potential
for $\sim$20 ps timing for charged particles \cite{Papaevangelou-MPGD2015-picosecond}.
It consists of a small amplification
gap (64 $\mu$m), a thin mesh (10 nm Al) coated with CsI to act as a reflective photocathode
or, in the semitrasparent version, a CsI photocathode window separated by a small gap
with voltage bias large enough to provide preamplification:
in the latter condition (Fig. \ref{fig:picosec-CsI}, left),
a single photon time resolution of 200 ps has been measured,
providing encouraging indications
for the possibility to achieve the desired time resolution goal.

An important element for a future extensive use of Cherenkov photon conversion in
harsh environments is a robust photocathode: intensive studies for alternatives
to standard CsI are ongoing on either protective layers (graphene shield enhanced
photocathodes \cite{Moody-2013-graphene} or different UV photocathode materials.
A promising candidate has recently been identified: hydrogenated rich-graphite
nano-diamond grains of $\sim$ 250 nm offer robustness and high QE
\cite{Valentini-2017-nandiamond} (see Fig. \ref{fig:picosec-CsI}, right)

\section{The quest for visible light gaseous PDs}

\par The detection of Cherenkov light in the visible range
brings several great advantages with respect to the far UV case: larger photon yield
thanks to the larger bandwidth, better angular resolution thanks to the reduced
chromaticity of the radiator, wider range of optical media with good transparency,
including silica aerogel and larger tolerance to contaminants.

\par Combining these advantages with the opportunities offered by
gaseous PDs (cost-effectiveness for large area, low material budget, magnetic
insensibilty) is extremely appealing and a large R\&D effort is being invested in
developing the visible-light gaseous PD technology.

\par
Due to the high chemical reactivity and the fragility of visible light photo-converters
(alkali-antimony, bi- and multi-alkali), their use in gaseous detectors
is a challenging task: most of the commonly used materials are incompatible
with these photo-converters, ion-induced secondary electron emission is copious
and aging by ion bombardment is quite fast. Coating photocathodes with protective
films seriously reduces the QE \cite{Peskov-1994-visible}.

\par
Different detector architectures have been investigated for visible light gaseous PDs,
from grid photomultipliers \cite{Edmends-1988-visible} to CP-based PDs
\cite{Peskov-1999-visible}.
A CP is a plate made of a bundle of fine glass capillaries fused together,
uniformly arranged in a two-dimensional array. Typically, a CP thickness is ranging
from 0.4 mm to 1 mm, and the capillary holes can be from 5 $\mu$m to 50 $\mu$m wide.
A CP used as a hole-type MPGD \cite{Sakurai-1996-CP}
has metallic electrodes deposited on both surfaces.
CPs have high temperature resistance and low out-gassing, making them suited for
sealed devices, as in vacuum-based Micro Channel Plate (MCP)-PMT application.
CPs have been used coupled to MM \cite{Vavra-2004-visible}, also
with the capillaries inclined \cite{Vavra-2005-inclined} to achieve further reduction
of the IBF. A recent
development of funnel glass CP
allows achieving 83\%
of surface area opening, and thus a higher photoelectron collection efficiency
for semitransparent photocathode PDs.
\cite{Sugiyama-2016-funnel}.


\par
GEM-based gaseous PDs with bialkali photocathodes have been built and tested in sealed mode
operation \cite{Balcerzyk-2003-GEM-visible} \cite{Chechik-2003-GEM-visible}, with an
ion-blocking gate \cite{Breskin-2005-GEM-MHSP} or incorporating a Micro-Hole and Strip Plate
(MHSP) into a Multi-GEM cascaded multiplier \cite{Lyashenko-2009-GEM-MHSP} or with
flipped reverse-bias MHSP \cite{Breskin-2010-flipped-MHSP}: in the last case an IBF rate of
3$\times$10$^{-4}$ was achieved for a gain of 10$^5$. The polyimide GEMs material seems
however to be incompatible with visible light photocathodes \cite{Tokanai-2009-GEM}.
\par
Good compatibility (no degradation after 2 y) is instead seen when using using MPGDs made of
Pyrex glass plates and treated by sand micro-blasting, coupled with a Micromegas
\cite{Sumiyoshi-2011-GEM-CP}.
A dedicated system to study the gaseous PD QE degradation \cite{Moriya-2013-QE} allowed
to measure for a semitransparent bialkali photocathode in a sealed gaseous PMT with
micro-blasted glass CP and a 0.9 atm. of Ne/CF$_4$ 90/10 a QE reduction of about 20\%
after an integrated charge of 0.4 $\mu$C/mm$^2$.
\par
Using two micromeshes with conical holes, produced by chemical etching, having
holes of different diameter (120 $\mu$m in the first and 80 $\mu$m in the second micromesh)
and common pitch of 250 $\mu$m, arranged in a staggered pattern, with a bialkali
photocathode and Ne/CF$_4$ 90/10 at 1 bar, a gain of 10$^4$ and an IBF of 6$\times$10$^{-4}$
have been obtained \cite{Tokanai-2014-2MM}.
\par
A recently explored alternative to overcome the IBF problem is the use of graphene,
which is a single layer of carbon atoms arranged in a honeycomb lattice and is the
thinnest material to date. A suspended graphene monolayer could be used as a micromesh
with no transparency for gas molecules, atoms and ions, and possibly large transparency
for electrons with kinetic energy of few eV \cite{Franchino-2016-graphene}: if
demonstrated to be effective and practically usable over large areas it could
represent the ideal solution for the IBF suppression and open the way to the
production of visible light gaseous PDs.

\section{Cryogenic applications}

\par
Attempts to obtain high, stable electron multiplication directly in noble gases and
liquids at cryogenic temperatures using open geometry gaseous multipliers (wires,
needles or microstrips) have not been very successful \cite{Policarpo-1995-LXe-microstrip}
\cite{Kim-2004-LAr-avalanches} until GEM-based detectors, called CRyogenic  Avalanche
Detectors (CRAD) \cite{Buzulutskov-2003-CRAD} were introduced.
With triple GEM detectors providing gains approaching 10$^4$ \cite{Bondar-2006-GEM-2phase-Ar-Xe}
the possibility to develop of a dual-phase CRAD equipped with CsI photocathode,
able to detect both the charge ionization signal and the primary scintillation
signals from liquid Ar (LAr) was demonstrated \cite{Bondar-2007-GEM-CsI-2phase}.


\par
The use of THGEMs, (also called large Electron Multipliers: LEMs)
was soon proposed \cite{Rubbia-2004-giant-LAr}
in view of giant LAr scintillation, Cherenkov and charge imaging experiment; THGEMs
were indeed tested to be as effective as GEMs \cite{Bondar-2008-THGEM-GEM-2phase}
for cryogenic applications \cite{Rubbia-2004-first-2phase-LEM} \cite{Cantini-2008-LEM-TPC-gain}.
\par
It has recently been shown \cite{Xie-2013-PTFE} that
THGEMs made of PTFE (Polytetrafluoroethylene) are particularly suited for cryogenic
low noise applications because of their radiopurity.
\par
A wide variety of gaseous PDs have been proposed and tested
\cite{Buzulutskov-2012-CRAD}: they can be divided into three classes:\\
- PDs performing photon conversion and charge multiplication in a gas
separated by a window from the pure noble liquid (or gas) of the cryogenic detector.\\
- PDs operated directly in the noble gas of a dual-phase cryogenic detector.\\
- PDs immersed in the cryogenic noble liquid.
\par
An example belonging to the first class is the Gaseous PhotoMultiplier (GPM)
\cite{Arazi-2015-liquid-hole} formed by a triple THGEM
(each having thickness = 0.4 mm, hole diameter = 0.4 mm, hole pitch = 0.8 mm, rim = 50 $\mu$m)
with active diameter of 100 mm, held 2 mm apart, with CsI on the first one,
operated in Ne/CH$_4$,
coupled through a UV window to a small dual-phase LXe TPC. It has recently demonstrated
the ability to record both single photon and massive electro-luminescence signals
in the same operating conditions, with maximal gain above 10$^5$, assuring high single
photon detection efficiency and stable operation.
\par
A member of the second class is the dual-phase CRAD prototype
\cite{Bondar-2015-2phase-electroluminescence}
consisting of four THGEMs (10 $\times$ 10 cm$^2$),
two horizontally immersed in the LAr at 48 mm distance, biased to form a drift
region in the liquid which covered the second THGEM by 4 mm, forming an electron emission region.
A double-THGEM assembly with the first THGEM acting as the anode was placed
18 mm above the liquid surface, to form an electro-luminescence gap and provide multiplication.
A matrix of Geiger Avalanche Photo-Diodes and cyogenic PMTs with wavelength shifters registered
the scintillation light. A systematic study of the proportional electro-luminescence
was performed. A recent measurement \cite{Bondar-2016-2phase-ArN2}
showed that a small ($\sim$50 ppm) N$_2$ doping level of the Ar enhances the CRAD
sensitivity to the proportional electro-luminescence signal.
\par
A representative of the third class is the Liquid Hole Multiplier
\cite{Breskin-2013-liquid-hole-multiplier}, proposed as a cascade
of THGEMs (or analogous hole multipliers) with CsI photocathodes deposited on
their surfaces, immersed in the noble liquid. Photoelectrons from primary scintillation
or ionization electrons are focused into the electrode holes and give origin to
electro-luminescence in the intense electric field in the liquid inside the holes;
the amplification of the UV photons in the cascaded THGEM structure could result
in detectable signals. Liquid Xe proportional electro-luminescence in THGEM holes
was indeed observed \cite{Arazi-2013-electroluminescence} and its large intensity
attributed to the possible presence of bubbles (local dual-phase conditions) near
the THGEM holes \cite{Arazi-2015-bubbles}, which have recently been directly observed
\cite{Arazi-2015-direct-bubbles}. The possibility to exploit this phenomenon in
large-volume local dual-phase liquid TPCs is fascinating.
\par
Recent reviews on liquid noble gas radiation detection \cite{Chepel-2013-LAr-Xe}
and on Gaseous and dual phase TPCs for rare processes
\cite{Gonzalez-Diaz-2017-TPC}
provide comprehensive overviews on these subjects.

\section{Scintillation imaging and other detectors}

Scintillation proportional counters \cite{Charpak-1987-scintillation} exploit the photon
emission process induced by electron-molecule collisions; this emission can be so copious
to allow the use of external optical imaging. 
\par
A two-dimensional detector with high spatial resolution can be established by incorporating
an imaging element, such as a charge-coupled device (CCD), detecting the scintillation
light produced during the charge amplification process.
For thermal neutron imaging
a triple GEM operating in $^3$He-CF$_4$ gas mixtures \cite{Fraga-2002-scintillation}
has exploited the CF$_4$ strong scintillation component in the visible (500 nm - 700 nm)
which matches the spectral sensitivity of CCD cameras; recently a glass GEM
combined with a mirror-lens-CCD system has been used for high-resolution
X-Ray 3D computed tomography \cite{Fujiwara-2017-G-GEM-CT}; coupled to
a micro-structured $^{10}$B foil \cite{Fujiwara-2016-neutron-imaging} it has been
used for neutron imaging.
Gaseous X-ray detectors can provide
good energy resolution and space resolution at the same time \cite{Veloso-2017-X-ray}.

\par
MPGD-based scintillating imaging detectors have extraordinary potential,
could find application in many different fields outside physics
research and help advancement in novel detector invention.

\section{Conclusions}

Gaseous PDs are the most effective approach to instrument large surfaces at affordable costs.
MPGD-based PDs overcome the limitations of open geometry gaseous PDs.
A large effort to refine and consolidate MPGD-based PD technologies is taking place.
Large hybrid THGEM-Micromegas PDs, covering 1.4 $m^2$ have been successfully operated
for the 2016 run on the COMPASS RICH-1 detector. 
Many promising developments are advancing in the UV and visible light detection,
and in the very dynamic fields of cryogenic detectors and scintillation light imaging. 
Projects in several domains, in particular PID, neutrino physics, DM search, medical
applications, X-Ray and neutron imaging, etc. are proposing the use of gaseous PDs
related technologies.
The field has a bright future: technology consolidation and new applications are taking
place, large scale projects are progressing, new techniques are being developed and new
ideas are proposed.



\section*{References}


\end{document}